# Forecasting Value-at-Risk with Time-Varying Variance, Skewness and Kurtosis in an Exponential Weighted Moving Average Framework

Alexandros Gabrielsen[a,1], Paolo Zagaglia[b,1], Axel Kirchner[c,1] and Zhuoshi Liu[d,1]

This version: June 6, 2012

**Abstract**
This paper provides an insight to the time-varying dynamics of the shape of the distribution of financial return series by proposing an exponential weighted moving average model that jointly estimates volatility, skewness and kurtosis over time using a modified form of the Gram-Charlier density in which skewness and kurtosis appear directly in the functional form of this density. In this setting *VaR* can be described as a function of the time-varying higher moments by applying the Cornish-Fisher expansion series of the first four moments. An evaluation of the predictive performance of the proposed model in the estimation of 1-day and 10-day *VaR* forecasts is performed in comparison with the historical simulation, filtered historical simulation and GARCH model. The adequacy of the *VaR* forecasts is evaluated under the unconditional, independence and conditional likelihood ratio tests as well as Basel II regulatory tests. The results presented have significant implications for risk management, trading and hedging activities as well as in the pricing of equity derivatives.

[1]Acknowledgements: The authors express their gratitude to Ying Hu from Sumitomo Mitsui Banking Corporate Europe for her valuable comments and suggestions.
[a] Sumitomo Mitsui Banking Corporation, London, U.K.
[b] Corresponding author: Department of Economics, University of Bologna, Italy; paolo.zagaglia@unibo.it.
[c] Deutsche Bank, London, U.K.
[d] Bank of England, London, U.K.
**Disclaimer**: The views expressed in this paper solely reflect the views of the authors and are not necessarily those of Sumitomo Mitsui Banking Corporation Europe, Deutsche Bank or Bank of England.

## 1. Introduction

The last few decades have seen a growing number of scholars and market participants being concerned with the precision of typical Value-at-Risk measures. Value-at-Risk is the maximum expected loss to occur for a given horizon and for a given probability. A key challenge in estimating accurate *VaR* confidence internals arises from the accurate estimation of the conditional distribution of financial return series. So far in the literature, many models have been put forward that capture some of the typical stylized facts of financial time series such as volatility clustering and pooling; that is the tendency of large changes to be followed by large changes - of either sign - and small changes to be followed by small changes (see Mandlebrot, 1963).

One of the early models employed in capturing volatility is the equally weighted moving average model. This framework assumes that the *N*-period historic estimate of variance is based on an equally weighted moving average of the *N*-past one-period squared returns. However, under this formulation all past squared returns that enter the moving average are equally weighted and this may lead to unrealistic estimates of volatility. In this respect the exponentially weighted moving average (EWMA) framework proposed by J.P Morgan's RiskMetrics$^{TM}$ assigns geometrically declining weights on past observations with the highest weight been attributed to the latest (i.e. more resent) observation. By assigning the highest weight to the latest observations and the least to the oldest the model is able to capture the dynamic features of volatility. Other approaches in this direction are the celebrated ARCH and GARCH model proposed by Engle (1982) and Bollerslev (1986) respectively. The former introduces the Autoregressive Conditional Heteroscedasticity (ARCH), which models the variance of a time series by conditioning it on the square of lagged disturbances and the latter generalizes the ARCH model by considering the lagged variance as an explanatory variable.[2]

Volatility, however, is only one of the distributional moments that can provide a stylized representation of returns. Empirical evidence has shown that the empirical distribution of financial series is likely to be skewed and fat-tailed[3] (see Mandlebrot 1963, Bollerslev, 1987, Campbell and Siddique, 1999 and 2000, Alizadeh and Gabrielsen, 2011, among others).

---

[2] The exponentially weighted moving average (EWMA) estimator has proven to be very effective at forecasting the volatility of returns over short horizons, and often has been found to provide superior *VaR* forecasts compared with GARCH models (see Baillie and DeGennaro, 1990; Bollerslev, Chou and Kroner, 1992; Boudoukh, Richardson and Whitelaw, 1997; Alexander and Leigh, 1997).
[3] Skewness is a measure of the asymmetry and kurtosis is a measure of the peakedness of a probability distribution.

Failing to account for the distributional characteristics of the return series will have serious implications in risk management and specifically in the estimation of Value-at-Risk (see Pedrosa and Roll, 1998, Bond, 2001, Burns, 2002, Angelidis *et al.* 2004 and 2007, Wilhelmsson 2009, Alizadeh and Gabrielsen 2011, among others), in pricing of derivates (see Heston and Nandi 2000, and Tahani 2006, among others), in trading and heding activities (see Kostika and Markellos, 2007, Apergis and Gabrielsen, 2011), in portfolio allocation (see Sun and Yan, 2003, Harvey *et al.* 2004 and Jondeau and Rockinger, 2006, among others). Although the standard EWMA estimator will be consistent when returns are not-normally distributed, it will be asymptotically inefficient since it places too much weight to extreme returns (see Guermat and Harris, 2002).

Guermat and Harris (2002) propose a general power EWMA model which is based on the Generalized Error Distribution (GED). They apply the model on daily return series of US, UK and Japan portfolios and find that their model is able to capture the fat-tailed nature of most returns series and estimate superior VaR forecasts compared with the standard EWMA formulation. Lin, Changchien and Chen (2006) and Liu, Wu and Lee (2007) apply a dynamic power EWMA that is able to capture the time-varying tail-fatness and volatilities of financial returns. They both apply it on well diversified equity portfolios and find that the model offers substantial improvements on capturing the dynamic distributional return characteristics, and can significantly enhance the estimation accuracy of portfolio *VaR*. Shyan-Rong, *et al.* (2010) compare the performance of a variety of models such as EWMA, Power EWMA, Dynamic Power EWMA, GARCH among others on six daily stock index returns (i.e. S&P500, Dow Jones Industrial Average, Nasdaq Composite, Taiwan Stock Index, Nikkei 255 and FTSE 100). They find that the dynamic power EWMA out-performs all other specifications in forecasting volatility. Although, the proposed formulations are able to capture the dynamic nature of returns by allowing a shape parameter to vary over time, little has been done to extend the exponential weighted moving average to account for time-varying higher moments.

The aim of this study is to investigate the nature and dynamics of the shape of the distribution of the returns overtime. We propose a formulation that jointly estimates time-varying volatility, skewness and kurtosis in an exponentially weighted moving average framework using the Gram-Charlier series expansion and allows each process to have its own decay factor. The method is based on the use of the Gram-Charlier density in which skewness and kurtosis appear directly in the functional form of the distribution, and for this reason, it is very simple to estimate the different decay factors using the maximum likelihood estimation approach. Furthermore, the forecast of VaR measures is performed with the application of the Cornish-Fisher expansion, which is used to estimate the quantile as a function of the time-varying volatility, skewness and kurtosis at a fixed confidence level.

We propose an evaluation of the predictive performance of the model for VaR forecasts. We compare against the historical simulation, filtered historical simulation and GARCH model. The adequacy of the VaR forecasts is evaluated under the tests for unconditional, independence, conditional likelihood ratio and under the Basel II regulatory tests. The results have significant implications for risk management, trading and hedging activities, as well as in the pricing of equity derivatives.

This paper is organized as follows section 2 describes a detailed description of the methodology, section 3 presents the data utilized and discusses the estimation results, while section 4 concludes.

## 2. Methodology

Given a series of stock market index prices $p_t$ and the corresponding rate of return $r_t$ is then defined as the continuously compounded return (in percent)

$$r_t = 100 \cdot [\ln(p_t) - \ln(p_{t-1})] \quad (1)$$

where the index $t$ denotes the daily closing observations and $t = 1,2,\ldots,T$. Furthermore, the sample period is comprised by an estimation (in-sample) period with $N$ observations $t = 1,2,\ldots,N$ and an evaluation (out-of-sample) period with $n$ observations $t = N+1,\ldots,T$.

The exponential weighted moving average proposed by J.P. Morgan's RiskMetrics$^{TM}$ for the series of returns $r_t$ is given as

$$r_t = \mu + \varepsilon_t, \quad \varepsilon_t \sim iid(0, \sigma_t) \quad (2)$$

$$\sigma_t^2 = \lambda \sigma_{t-1}^2 + (1-\lambda) \varepsilon_{t-1}^2 \quad (3)$$

where $\lambda \ (0 < \lambda < 1)$ denotes the decay factor, $r_t$ the returns, $\varepsilon_t$ the innovation terms and $\sigma_t^2$ denotes the variance at time $t$.

In order to model the dynamics of skewness and kurtosis the modified exponential weighted average is formalised in the following way

$$r_t = \mu + \varepsilon_t, \quad \varepsilon_t = \eta_t \sigma_t, \quad \eta_t \sim iid(f(\eta_t)) \tag{4}$$

$$\sigma_t^2 = \lambda_1 \sigma_{t-1}^2 + (1-\lambda_1)\varepsilon_{t-1}^2 \tag{5}$$

$$s_t^3 = \lambda_2 s_{t-1}^3 + (1-\lambda_2)\eta_{t-1}^3 \tag{6}$$

$$k_t^4 = \lambda_3 k_{t-1}^4 + (1-\lambda_3)\eta_{t-1}^4 \tag{7}$$

where $\lambda_i$, $0 < \lambda_i < 1$ which denotes the decay factor for each specification, $r_t$ the returns, $\varepsilon_t$ the error terms, $\eta_t$ the standardized error terms, $\sigma_t^2$ denotes the variance at time $t$, $s_t^3$ denotes the skewness at time $t$, and $k_t^4$ the kurtosis at time $t$. This formulation assumes that the standardized residuals of the return series follow a Gram-Charlier distribution. The Gram-Charlier Type A distribution is an approximate probability density function of the normal density function in terms of the Hermite polynomials and it is estimated as follows

$$f(\eta_t) = \phi(\eta_t) g(\eta_t) \tag{8}$$

where $\phi(\cdot)$ denotes the standard normal density with zero mean and unit variance, $g(\cdot)$ is a polynomial function that matches the first moments of the standardized residual's probability density function and is represented as

$$g(\eta_t) = \sum_{i=0}^{n} c_i He_i(\eta_t) \tag{9}$$

$$He_i(\eta_t) = (-1)^i \frac{\partial^i \phi}{\partial \eta_t^i} \frac{1}{\phi(\eta_t)} \tag{10}$$

where $He_i(\cdot)$ denote the Hermite polynomials and when truncating at the fourth moment Equations 9 and 10 become

$$g(\eta_t) = 1 + \frac{s_t}{6} He_3(\eta_t) + \frac{k_t - 3}{24} He_4(\eta_t) \tag{11}$$

$$He_3(\eta_t) = \eta_t^3 - 3\eta_t \tag{12}$$

$$He_4(\eta_t) = \eta_t^4 - 6\eta_t^2 + 3 \tag{13}$$

Finally the Gram-Charlier density assumes the following form

$$f(\eta_t) = \phi(\eta_t)\left[1 + \frac{s_t}{3!}(\eta_t^3 - 3\eta_t) + \frac{k_t - 3}{4!}(\eta_t^4 - 6\eta_t^2 + 3)\right] \tag{14}$$

The problem with function 14 is that it is not really a density function since, for some parameter values of skewness and kurtosis, it can become negative and thus the integral of $f(\cdot)$ may not be equal to one. In order to obtain a well-defined positive density function, Galland and Tauchen (1989) describe the density in terms of the square expansion terms $g(\cdot)$ and divide it by the function $h(\cdot)$, which denotes the integral of $f(\cdot)$.[4] The density therefore is defined as

$$f(\eta_t) = \frac{\phi(\eta_t) g^2(\eta_t)}{h(\eta_t)} \quad (15)$$

where

$$h(\eta_t) = 1 + \frac{s_t^2}{3!} + \frac{(k-3)^2}{4!} \quad (16)$$

We should stress that the Gram-Charlier series expansion nests the Gaussian distribution when $s_t = 0$ and $k_t = 3$.

The estimation of the model parameters is obtained by maximising the likelihood function. This is based on the assumption that the residuals follow a Gram-Charlier density. The log-likelihood function for one observation can then be written as

$$l_t = -\frac{1}{2}\log(2\pi) - \frac{1}{2}\log(\sigma_t^2) - \frac{1}{2}\eta_t^2 + \log[g^2(\eta_t)] - \log(h(\eta_t)) \quad (17)$$

In the empirical application, we compare the performance of our model in forecasting VaR against the nominal GARCH(1,1) model proposed by Bollerslev (1986):

$$r_t = \mu + \varepsilon_t, \quad \varepsilon_t \sim iid(0, \sigma_t) \quad (18)$$

$$\sigma_t^2 = a_0 + a_1 \varepsilon_{t-1}^2 + a_2 \sigma_{t-1}^2 \quad (19)$$

where $r_t$ denotes the returns, $\varepsilon_t$ the error terms and $\sigma_t^2$ the conditional variance at time $t$. The estimation of the parameters of the GARCH(1,1) is undertaken by the maximization of the empirical likelihood function. For one observation, this function takes the form:

$$l_t = -\frac{1}{2}\log(2\pi) - \frac{1}{2}\log(\sigma_t^2) - \frac{\varepsilon_t^2}{2\sigma_t^2} \quad (20)$$

We use Matlab routines to estimate jointly all the parameter values using the Broyden-Fletcher-Goldfarb-Shanno (BFGS) quasi-Newton optimization algorithm for the numerical maximisation of the log-likelihood functions.

## 3. Evaluation of Value-at-Risk Estimates

Forecast evaluation is one of the most important aspects of any forecasting exercise and especially in the evaluation of accurate Value-at-Risk estimates. Value-at-Risk is a measure of the market risk of a portfolio and refers to the particular amount of money that is likely to be lost due to market fluctuations over a period of time and for given a probability. The VaR at time *t* at *α%* significance level is estimated as

$$VaR_{t+n} = \mu_{t+n} - F^{-1}(\alpha)\hat{\sigma}_{t+n} \qquad (21)$$

where $F^{-1}(\alpha)$ denotes the empirical quantile of assumed distribution function, *n* is the forecasted horizon and $\hat{\sigma}_{t+n/t}$ is the *t+n* volatility forecast. Traditional quantitative risk models assume that financial return series are normally distributed. However, empirical evidence has shown that the empirical distribution of returns is often skewed, fat-tailed and peaked around the mean. When these aspects are ignored, the calculation of VaR is seriously compromised (see Pedrosa and Roll, 1998; and Alizadeh and Gabrielsen, 2012). Therefore, in order to incorporate the dynamics of higher-moments we apply the Cornish-Fisher expansion to approximate the inverse cumulative density function. The Cornish-Fisher expansion can be viewed as an expansion of the Gaussian Normal density function augmented with terms that capture the dynamic nature of skewness and kurtosis and can be formalised as follows

$$F^{-1}(\alpha) = \varphi^{-1}_{1-a}\left\{1 + \frac{s_t}{3!}\left[\left(\varphi^{-1}_{1-a}\right)^2 - 1\right] + \frac{k_t}{4!}\left[\left(\varphi^{-1}_{1-a}\right)^3 - 3\varphi^{-1}_{1-a}\right]\right\} \qquad (22)$$

where $\varphi(\cdot)^{-1}$ denotes the inverse cumulative density function of the standard normal distribution and $s_t$ and $k_t$ the skewness and kurtosis estimates from the modified exponential weighted-average model.

---

[4] A proof that the Galland and Tauchen (1989) density is a true density is presented in Appendix A.

We also evaluate the performance of the Historical Simulation (HS) and Filtered Historical Simulation (FHS). The Historical Simulation uses past returns to estimate the cumulative distribution function, hence taking into consideration asymmetries and fat tails. The Historical Simulation is defined as:

$$VaR_t = (F_a)\left(\{r_{t-i}\}_{i=t-i-N}^{t-1}\right) \qquad (23)$$

where the right hand of the equation defines the *a* percentile of *N* past returns. An extension of the historical simulation that assumes that returns are independent and identically distributed is represented by the Filtered Historical Simulation. This is defined as:

$$VaR_t = F_a\left(\{z_{t-i}\}_{i=t-i-N}^{t-1} \mid \theta\right)\sigma_t \qquad (24)$$

where $z_{t-i}$ are the standardized residuals and $\sigma_t$ is the standard deviation of the returns.

The adequacy of the VaR estimates is examined over a back-testing exercise, where actual profits and losses are compared to the corresponding Value-at-Risk forecasts of the various models. Regulators currently employ three techniques to evaluate the adequacy of the VaR models: the binomial, the interval forecast and distribution forecast methods.

The time until first failure test (TUFF) is based on the number of observations before the first exception (see Kupiec, 1995). The null hypothesis is, $H_0 : a = \hat{a}$ and the corresponding LR test is

$$LR_{TUFF} = -2\ln\left[\hat{a}(1-\hat{a})^{n-1}\right] + 2\ln\left[\frac{1}{n}(1-n)^{n-1}\right] \sim \chi^2(1) \tag{25}$$

where $n$ denotes the number of observations before the first exception. The $LR_{TUFF}$ is asymptotically distributed as a $\chi^2(1)$. Kupiec (1995) argues that the test has limited power to distinguish among alternative hypothesis since all observations after the first exception are ignored.

Christoffersen (1998) develops an interval forecast method that examines whether *VaR* estimates exhibit correct coverage. Christoffersen (1998) emphasizes the importance of conditional testing, which takes into account not only the frequency of VaR violations but also the timing of occurrence, which measures the clustering of failures. Christoffersen (1998) approach can be separated into the unconditional coverage, the independence and the conditional coverage tests. Therefore, the rejection of a model can be categorized as the unconditional coverage failure or the exception clustering, or both.

Given a time series of past ex-ante VaR forecasts, and ex-post returns, *r*, a hit sequence or indicator function can be estimated as

$$I_{t+i} = \begin{cases} 1, & r_{t+i} < -VaR_{t+i}^p \\ 0, & r_{t+i} > -VaR_{t+i}^p \end{cases}, \quad i = 1,...,T \tag{26}$$

the indicator faction returns one if the loss is larger than the estimated *VaR* and zero otherwise. The VaR model is said to be efficient if the indicator function is independently distributed over time as a Bernoulli variable.

The unconditional coverage examines whether the estimated α% VaR violations fall within the theoretical number of α% VaR violations:

$$LR_{UC} = -2\log\left[\frac{p^{n_1}(1-p)^{n_0}}{\hat{\pi}^{n_1}(1-\hat{\pi})^{n_0}}\right] \sim \chi^2(1) \tag{27}$$

where $n_1$ the number of 1's in the indicator series, $n_0$ the number of 0's in the indicator series and $\hat{\pi} = n_1/(n_1 + n_0)$.

The test of independence tests for the clustering of VaR exceptions under the hypothesis of an independently distributed failure process against the alternative hypothesis of first order Markov failure process. The likelihood ratio test is

$$LR_{IND} = -2\log\left[\frac{(1-\hat{\pi}_2)^{n_{00}+n_{10}}(1-\hat{\pi}_2)^{n_{01}+n_{11}}}{(1-\hat{\pi}_{01})^{n_{00}}\hat{\pi}_{01}^{n_{01}}(1-\hat{\pi}_{11})^{n_{10}}\hat{\pi}_{11}^{n_{11}}}\right] \sim \chi^2(1) \qquad (28)$$

where $n_{ij}$ is the number of $i$ values followed by $j$ value in the indicator function, $\pi_{ij} = \Pr\{I_t = i / I_{t-1} = j\}$, $\hat{\pi}_{01} = n_{01}/(n_{00}+n_{01})$, $\hat{\pi}_{11} = n_{11}/(n_{10}+n_{11})$ and $\hat{\pi}_2 = (n_{01}+n_{11})/(n_{00}+n_{01}+n_{10}+n_{11})$. In the special case when $\hat{\pi}_{11} = 0$ then the independence test can be computed as

$$LR_{IND} = (1-\hat{\pi}_{01})^{n_{00}}\hat{\pi}_{01}^{n_{01}} \sim \chi^2(1) \qquad (29)$$

Finally, the correct conditional coverage jointly tests for independence and correct coverage, with the test statistics as:

$$LR_{CC} = LR_{UC} + LR_{IND} \sim \chi^2(2) \qquad (30)$$

The regulatory guidelines prescribed by the 1996 amendment to the 1988 Basel Accord require commercial banks in the G-10 countries to carry out standardised back-tests that define the capital adequacy standards. The capital requirements are, therefore, depended on both the portfolio risk and the back-testing outcome of the bank's internal VaR model. According to Campbell (2005), the capital requirements are set as the larger of either the bank's current assessment of the 1% VaR over the following 10 trading days, or as a multiple of the bank's average reported 1% VaR over the previous 60 trading days, plus an additional amount that reflects the underlying credit risk of the bank's portfolio. This amount is computed as:

$$MRC_t = \max\left[VaR_t(0.01), S_t \frac{1}{60}\sum_{i=0}^{59} VaR_{t-i}(0.01)\right] + c \qquad (31)$$

where $S_t$ denotes a factor that multiplies the average of previously reported VaR estimates. This factor is determined by classifying the number of 1% *VaR* violations in the previous 250 trading days *x* into three categories

$$S_t = \begin{cases} 3, & x \leq 4, \quad green \\ 3 + 0.2(x-4), & 5 \leq x \leq 9, \quad yellow \\ 4, & x \succ 10, \quad red \end{cases} \qquad (32)$$

This expression highlights that as the number of violations increase so does the multiplication factors that determines the market risk capital. For example a model is classified as green when there is more than 99.99% probability that it's estimated 1% VaR violations fall within the theoretical (1%) number of VaR violations.

The models deemed adequate are those that generate a coverage rate less than the nominal, and that are able to pass both the conditional and unconditional coverage tests.

## 4. Data Description and Empirical Results

The data comprising the study is daily prices of the S&P 500, NASDAQ, FTSE 100, DAX 30 and CAC 40 equity indices for the period from 02/01/1992, 02/01/1992, 17/01/1991, 07/02/1992 and 11/02/1991 to 30/06/2011 respectively. The data is readily available from Datastream and non-trading days have been removed in order to avoid downsize bias. The descriptive statistics for the returns of the equity indices are reported in Table 1.The coefficients of skewness are negative for the returns of the equity indices, signifying a bias towards downside exposure. This contrasts sharply with positive skewness, which indicates the possibility of large positive returns (see Campbell and Siddique, 2000). The coefficients of excess kurtosis are above three indicating the distribution of the returns is leptokurtic; which means that the distribution has acute peakedness and fatter tails. The largest coefficient of excess kurtosis is reported for the S&P 500 followed by the FTSE 100 index, and highlights that these indices account for larger deviations in their returns. Finally, the Jargue-Bera test reveals significant departures from normality for all series at a 1% significance level for all indices.

### 4.1. In-Sample Analysis

The in-sample period is for S&P 500 from 2$^{nd}$ January 1992 to 7$^{th}$ July 2009, for NASDAQ s from 2$^{nd}$ January 1992 to 7$^{th}$ July 2009, FTSE 100 from 17$^{th}$ January 1991 to 7$^{th}$ July 2009, for DAX 30 is from 7$^{th}$ February 1992 to 23th June 2009, and CAC 40 is from 11$^{th}$ February 1991 to 23$^{th}$ June 2009. The first model to be examined is RiskMetrics$^{TM}$ based on a decay factor of 0.94. The unconditional volatility for the various equity indices for the in-sample period is presented in Figure 2. It is observed that the period between 1993 to mid 1997 and 2004 to 2007 is characterized by a low volatility period, whereas from 1998 to 2003 and 2008 to 2011 which are the dot com and credit crisis periods, a higher volatility period is observed.

The estimated decay factors for the volatility, skewness and kurtosis processes for the EWMA-SK model are presented in Table 2 and are significant for all models**.** A consistent pattern across the decay factors is observed. That is the decay factor for the volatility process has increased compared with the decay factor employed by RiskMetrics$^{TM}$ and overall is larger than the decay factors for skewness and kurtosis. The decay factors for the volatility process ranges between 0.935 for the DAX 30 to 0.980 for the NASDAQ, while the decay factors for the skewness process ranges between 0.948 for the DAX 30 and 0.969 for the S&P 500 and for the kurtosis process between 0.925 for the S&P 500 and 0.954 for the NASDAQ. The in-sample volatility, skewness and kurtosis are presented in Figure 3. It is observed that time-varying skewness and kurtosis fluctuate more and exhibit large spikes – negative spikes for time-varying skewness - during periods of high volatility. This means that the negative spikes in the time-varying skewness and positive spikes for the time-varying kurtosis highlight sharp deteriorating changes in the market conditions. Therefore, the extreme movements captured by the dynamics of skewness and kurtosis may have serious implications for risk management and, especially, in the estimation of VaR. Similar results are presented in Leon, *et al.,* (2005), Alizadeh and Gabrielsen (2012) and Apergis and Gabrielsen (2012).

The estimated parameters for the GARCH-N model, which we consider as a benchmark model, are reported in Table 3. The coefficients of the lagged squared error, $\beta_1$, and lagged conditional variance, $\beta_2$, are significant in all models. The values of the $\beta_1$ coefficient range between 0.092 for the DAX 30 to 0.070 for the S&P 500. For the $\beta_2$ coefficient, they vary between 0.899 for the DAX 30 and 0.922 of the CAC 40 index. Finally, Figure 4 presents the conditional variance for the in-sample period for the various models.

### 4.2. Out-of-Sample Evaluation

The performance of the VaR models is evaluated using a process known as back-testing. The back-testing exercise is undertaken for 1% VaR forecasts[5] over the period 7$^{th}$ July 2009 to 29$^{th}$ June 2011 for the S&P 500, NASDAQ and FTSE 100, 22$^{th}$ June 2009 to 29$^{th}$ June 2011 for the DAX 30 index and 23$^{th}$ June 2009 to 29$^{th}$ June 2011 for the CAC 40 index, with a total of 500 observations. The metrics employed are the Percentage of Failures (%), Christoffersen (1998) unconditional coverage, independence, and conditional coverage log-likelihood tests, along with Basel II test and are presented in Table 4. The Percentage of Failures (%) illustrates that the RiskMetrics, EWMA-SK and GARCH-N tend to exhibit the lowest on average percentage for both 1-day and 2-week VaR forecasts. For example under the FTSE 100 index and 1-day horizon the GARCH-N exhibits the lowest failures (0.20%), followed by RiskMetrics and EWMA-SK (0.41%) while HS and FHS exhibit PF above 1%; whereas for the 2-week horizon EWMA-SK exhibits the lowest PF (0.80%) followed by GARCH-N, HS, RiskMetrics and FHS.

---

[5] The 1% confidence level is selected as it is the level suggested by Basel II.

The likelihood ratio test for the unconditional coverage over the back-testing period is rejected at a 5% significance level for the RiskMetrics, EWMA-SK and GARCH-N for the S&P 500 index, the FHS, RiskMetrics, EWMA-SK and GARCH-N for the NASDAQ index, the HS, FHS and GARCH-N for the FTSE 100 index for 1-day horizon. Similarly for 10-day horizon the unconditional coverage is rejected at a 5% significance level for the HS, FHS and EWMA-SK for the S&P 500 index, the HS, FHS, RiskMetrics and EWMA-SK for the NASDAQ index, the HS and FHS for the DAX 30 and CAC 40 index. The results of the likelihood ratio test of independence indicate that the majority of models do not exhibit clustering of violations with the exception of the HS, FHS, RiskMetrics and EWMA-SK models for the FTSE 100 index and 1-day VaR the HS, FHS for the S&P 500, NASDAQ, DAX 30 and CAC 40 indices for the 10-day VaR along with RiskMetrics and EWMA-SK for the DAX 30 and CAC 40 indices for the 10-day horizon. These results indicate that not all models are to be relied for longer horizon estimates due to of clustering of violations. The likelihood ratio test of the conditional coverage is rejected at a 5% significant level for 1-day VaR for the RiskMetrics and EWMA-SK for the S&P 500 index, the RiskMetrics and GARCH-N for the NASDAQ index, the HS, FHS RiskMetrics and EWMA-SK for the FTSE 100 index. Moreover, the test is rejected at a 5% significant level over the 10-day horizon for the HS and FHS for the S&P 500, NASDAQ, DAX 30 and CAC 40 indices, along with the RiskMetrics for the DAX 30 index and EWMA-SK for the DAX 30 and CAC 40 indices. The VaR estimates of the GARCH-N perform well for the longer horizon, whereas the EWMA-SK for the short horizon. Similar results are presented in Guermat and Harris (2002) and Changchien and Chen (2006).

The models are also compared with respect to compliance with Basel II test which groups models into three categories: green, yellow and red depending on the number 1% VaR violations. For 1-day VaR the EWMA-SK and GARCH-N for the FTSE 100 and DAX 30 along with RiskMetrics for the FTSE 100 index fall within the green zone. In the yellow zone lay the HS, FHS, EWMA-SK and GARCH-N of the S&P 500, NASDAQ and CAC 40 indices including the HS, FHS and RiskMetrics for the DAX 30 index. In the red zone are the RiskMetrics of the S&P 500 and NASDAQ indices, as well as the HS and FHS of the FTSE 100 index. For the 2-week VaR EWMA-SK and GARCH-N fall in the green zone for all indices, followed by RiskMetrics with only FTSE 100 being in the yellow zone and finally by the HS and FHS which fall between the yellow and red rejection regions for all indices. This test revealed both the EWMA-SK and GARCH-N models performed well for both 1-day and 2-week horizons.

Figures 5 to 9 present the 1-day and 2-week VaR estimates HS, FHS, RiskMetrics, EWMA-SK and GARCH-N models over the back-testing period. The VaR forecasts for the EWMA-SK model appear to behave more erratic compared with RiskMetrics. This occurs because the *VaR* of the EWMA-SK contains estimates of the forecasted skewness and kurtosis, which over the examined period they increase significantly and exhibit spike thus affecting the estimation of the Cornish-Fisher approximation.

Concluding, the back-testing application delivers mixed results in terms of model validation. This outcome is not uncommon in this kind of studies (e.g., see Marcucci, 2009 and Alizadeh and Gabrielsen, 2012). A possible explanation is that the back-testing period was one of the most turbulent periods; which in turn translates into more erratic higher moment forecasts and hence impacting the VaR estimates of the EWMA-SK model. However, it is important to note that the EWMA-SK performs on average as well as the GARCH model for both horizon periods, and out-performs the RiskMetrics; compared with the studies Guermat and Harris (2002) and Changchien and Chen (2006) who find that GARCH models out-perform RiskMetrics$^{TM}$ for short and long horizons in forecasting Value-at-Risk.

## 5. Conclusions

The aim of this study was to propose a formulation that jointly estimates time-varying volatility, skewness and kurtosis in an exponentially weighted moving-average framework using the Gram-Charlier series expansion, and allows each process to have its own decay factor. A maximum likelihood estimation approach is used to estimate the decay factor for volatility, skewness and kurtosis. It was observed that the decay factor for the volatility process has increased compared with the decay factor used by RiskMetrics$^{TM}$ for daily return series, and overall is larger compared with the decay factors for skewness and kurtosis. Furthermore, we observe that, during deteriorating economic conditions, time-varying skewness exhibits negative spikes (i.e. bias towards down-size bias), while time-varying kurtosis displays positive spikes. This suggests that higher moments are able to identify and capture the dynamic characteristics of the various return series (see also Apergis and Gabrielsen, 2012, and Alizaden and Gabrielsen, 2012).

The performance of the proposed model along with RiskMetrics$^{TM}$, GARCH-N, HS and FHS is compared for 1-day and 2-week VaR forecasts at 1% confidence level. The adequacy of the VaR estimates was evaluated using the Christoffersen (1998) back-testing procedure and Basel II regulatory test. The results from the model validation process were mixed and may be due to the selection of the back-testing period, which coincided with the recent credit crisis period. The results are in line with the Marcucci (2009) and Alizadeh and Gabrielsen (2012) who do not find a uniformly accurate model. However, it is highlighted that EWMA-SK performs on average as well as the GARCH model for both horizon periods, and out-performs the RiskMetrics; which reflects a result different from other studies which find that GARCH models out-perform RiskMetrics$^{TM}$ for short and long horizons in forecasting Value-at-Risk.

**Table 1 Descriptive Statistics of the Returns**

| Statistics | S&P500 | NASDAQ | FTSE 100 | DAX 30 | CAC 40 |
|---|---|---|---|---|---|
| Mean | 0.017 | 0.025 | 0.011 | 0.026 | 0.013 |
| Standard Deviation | 1.199 | 1.640 | 1.178 | 1.490 | 1.434 |
| Minimum | -9.470 | -10.168 | -9.265 | -7.433 | -9.472 |
| Maximum | 10.957 | 13.255 | 9.384 | 10.797 | 10.595 |
| Skewness | -0.201 | -0.027 | -0.101 | -0.051 | -0.026 |
| Kurtosis | 12.482 | 8.555 | 9.649 | 7.602 | 7.748 |
| Jarque-Bera[1] | 16560 | 5675 | 8136 | 3896 | 4146 |

**Notes:**
The table presents the descriptive statistics of the return series. The sample period for S&P 500, NASDAQ, FTSE 100, DAX 30 and CAC 40 indices are from 02/01/1992, 02/01/1992, 17/01/1991, 07/02/1992 and 11/02/1991 to 30/06/2011 respectively. The Jarque-Bera (1980) test, tests for departure from normality and is chi-square asymptotic with two degrees-of-freedom, the 5% and 1% statistics are 5.99 and 9.21 respectively. It's statistic is defined in terms of the number of observations, n, sample skewness s, and sample kurtosis, k and it is described as: $JB = \frac{n}{6}\left(s + \frac{k-3}{4}\right)$

**Table 2 EWMA-SK Parameter Estimation**

|  | S&P500 | NASDAQ | FTSE 100 | DAX 30 | CAC 40 |
|---|---|---|---|---|---|
| $\lambda_1$ | 0.973 | 0.980 | 0.969 | 0.935 | 0.968 |
|  | (5012.272) | (10491.883) | (13804.854) | (2104.146) | (2154.760) |
| $\lambda_2$ | 0.969 | 0.966 | 0.961 | 0.948 | 0.948 |
|  | (2004.967) | (2577.322) | (2574.322) | (1595.139) | (2009.322) |
| $\lambda_3$ | 0.929 | 0.954 | 0.943 | 0.942 | 0.942 |
|  | (1516.674) | (3105.102) | (2372.410) | (2467.714) | (1909.631) |

**Notes:**
The table presents the estimation results for the parameters of the as well in parentheses are the t-statistics for the EWMA-SK model. The estimation is performed by the method of quasi maximum likelihood using the BFGS algorithm in Matlab 7.12 software package The sample period for S&P 500 is from 2nd January 1992 to 7th July 2009, for NASDAQ s from 2nd January 1992 to 7th July 2009, FTSE 100 is from 17th January 1991 to 7th July 2009, for DAX 30 is from 7th February 1992 to 23th June 2009, and CAC 40 is from 11th February 1991 to 23th June 2009.

**Table 3 GARCH-N Parameter Estimation**

|       | S&P500    | NASDAQ    | FTSE 100  | DAX 30    | CAC 40    |
|-------|-----------|-----------|-----------|-----------|-----------|
| $a_0$ | 0.047     | 0.070     | 0.040     | 0.069     | 0.050     |
|       | (3.883)   | (4.346)   | (3.231)   | (4.304)   | (2.977)   |
| $\beta_0$ | 0.007 | 0.013     | 0.010     | 0.021     | 0.015     |
|       | (6.777)   | (6.204)   | (5.131)   | (7.385)   | (4.602)   |
| $\beta_1$ | 0.070 | 0.080     | 0.086     | 0.092     | 0.071     |
|       | (14.015)  | (15.111)  | (12.812)  | (13.001)  | (12.103)  |
| $\beta_2$ | 0.925 | 0.915     | 0.907     | 0.899     | 0.922     |
|       | (174.231) | (159.390) | (128.642) | (122.589) | (140.482) |

**Notes:**
The table presents the estimation results for the parameters of the as well in parentheses are the t-statistics for the GARCH model with Gaussian Normal innovation terms. The estimation is performed by the method of quasi maximum likelihood using the BFGS algorithm in Matlab 7.12 software package The sample period for S&P 500 is from 2$^{nd}$ January 1992 to 7$^{th}$ July 2009, for NASDAQ s from 2$^{nd}$ January 1992 to 7$^{th}$ July 2009, FTSE 100 is from 17$^{th}$ January 1991 to 7$^{th}$ July 2009, for DAX 30 is from 7$^{th}$ February 1992 to 23th June 2009, and CAC 40 is from 11$^{th}$ February 1991 to 23$^{th}$ June 2009.

**Table 4 Value-at-Risk Back-Testing**

| | Percentage of Failures (%) | | | | | | | | | |
|---|---|---|---|---|---|---|---|---|---|---|
| | **1-day** | | | | | **10-day** | | | | |
| | *HS* | *FHS* | *RiskMetrics* | *EWMA-SK* | *GARCH-N* | *HS* | *FHS* | *RiskMetrics* | *EWMA-SK* | *GARCH-N* |
| *S&P500* | 1.80% | 1.80% | 2.80% | 2.20% | 2.20% | 2.65% | 3.26% | 0.00% | 0.20% | 0.00% |
| *NASDAQ* | 1.80% | 2.20% | 3.00% | 2.20% | 2.60% | 2.44% | 2.44% | 0.20% | 0.20% | 0.00% |
| *FTSE100* | 3.26% | 3.46% | 0.41% | 0.41% | 0.20% | 1.40% | 1.60% | 1.60% | 0.80% | 1.40% |
| *DAX 30* | 1.80% | 1.80% | 1.80% | 1.40% | 1.20% | 2.85% | 3.05% | 0.81% | 0.41% | 0.20% |
| *CAC 40* | 1.80% | 2.00% | 1.80% | 1.60% | 1.80% | 3.67% | 3.67% | 0.81% | 0.61% | 0.41% |
| | **Likelihood Ratio of Unconditional Coverage** | | | | | | | | | |
| | **1-day** | | | | | **10-day** | | | | |
| | *HS* | *FHS* | *RiskMetrics* | *EWMA-SK* | *GARCH-N* | *HS* | *FHS* | *RiskMetrics* | *EWMA-SK* | *GARCH-N* |
| *S&P500* | 2.613 | 2.613 | 10.994 | 5.419 | 5.419 | 9.271 | 15.877 | 0.000 | 4.669 | 0.000 |
| *NASDAQ* | 2.613 | 5.419 | 13.162 | 5.419 | 8.973 | 7.371 | 7.371 | 4.669 | 4.669 | 0.000 |
| *FTSE100* | 15.877 | 18.349 | 2.245 | 2.245 | 4.669 | 0.719 | 1.538 | 1.538 | 0.217 | 0.719 |
| *DAX 30* | 2.613 | 2.613 | 2.613 | 0.719 | 0.190 | 11.329 | 13.534 | 0.182 | 2.245 | 4.669 |
| *CAC 40* | 2.613 | 3.914 | 2.613 | 1.538 | 2.613 | 20.943 | 20.943 | 0.182 | 0.872 | 2.245 |
| | **Likelihood Ratio of Independence** | | | | | | | | | |
| | **1-day** | | | | | **10-day** | | | | |
| | *HS* | *FHS* | *RiskMetrics* | *EWMA-SK* | *GARCH-N* | *HS* | *FHS* | *RiskMetrics* | *EWMA-SK* | *GARCH-N* |
| *S&P500* | 0.000 | 0.000 | 0.710 | 1.429 | 0.000 | 14.675 | 16.560 | 0.000 | 0.001 | 0.000 |
| *NASDAQ* | 0.000 | 0.000 | 0.000 | 0.000 | 0.000 | 16.147 | 9.903 | 0.001 | 0.001 | 0.000 |
| *FTSE100* | 37.617 | 35.034 | 8.846 | 8.846 | 0.001 | 0.000 | 0.000 | 0.000 | 0.000 | 0.000 |
| *DAX 30* | 0.000 | 0.000 | 0.000 | 0.000 | 0.000 | 13.349 | 18.029 | 14.923 | 8.846 | 0.001 |
| *CAC 40* | 0.000 | 0.000 | 2.126 | 2.566 | 2.126 | 32.671 | 25.777 | 5.426 | 6.765 | 0.000 |

| | Likelihood Ratio of Conditional Coverage | | | | | | | | | |
|---|---|---|---|---|---|---|---|---|---|---|
| | 1-day | | | | | 10-day | | | | |
| | HS | FHS | RiskMetrics | EWMA-SK | GARCH-N | HS | FHS | RiskMetrics | EWMA-SK | GARCH-N |
| SP500 | 2.613 | 2.613 | 11.704 | 6.848 | 5.419 | 23.946 | 32.437 | 0.000 | 4.670 | 0.000 |
| NASDAQ | 2.613 | 5.419 | 13.162 | 5.419 | 8.973 | 23.518 | 17.274 | 4.670 | 4.670 | 0.000 |
| FTSE 100 | 53.494 | 53.383 | 11.091 | 11.091 | 4.670 | 0.719 | 1.538 | 1.538 | 0.217 | 0.719 |
| DAX 30 | 2.613 | 2.613 | 2.613 | 0.719 | 0.190 | 24.678 | 31.563 | 15.105 | 11.091 | 4.670 |
| CAC 40 | 2.613 | 3.914 | 4.739 | 4.104 | 4.739 | 53.614 | 46.721 | 5.608 | 7.637 | 2.245 |
| | BASEL II | | | | | | | | | |
| | 1-day | | | | | 10-day | | | | |
| | HS | FHS | RiskMetrics | EWMA-SK | GARCH-N | HS | FHS | RiskMetrics | EWMA-SK | GARCH-N |
| S&P500 | Yellow | Yellow | Red | Yellow | Yellow | Yellow | Red | Green | Green | Green |
| NASDAQ | Yellow | Yellow | Red | Yellow | Yellow | Yellow | Yellow | Green | Green | Green |
| FTSE 100 | Red | Red | Green | Green | Green | Green | Yellow | Yellow | Green | Green |
| DAX 30 | Yellow | Yellow | Yellow | Green | Green | Red | Red | Green | Green | Green |
| CAC 40 | Yellow | Yellow | Yellow | Yellow | Yellow | Red | Red | Green | Green | Green |

Notes:
The table reposts the percentage of failure; Christoffersen's (1998) likelihood ratio tests and BASEL II model categorization. The likelihood ratio test of the unconditional coverage and independenceare are distributed as chi-square asymptotic with one degrees-of-freedom. The 1% and 5% critical values for $\chi^2(1)$ are 6.634 and 3.841. The likelihood ratio test of the conditional coverage is chi-square asymptotic with two degrees-of-freedom. The 1% and 5% critical values for $\chi^2(1)$ are 9.21 and 5.99. The back-testing period is for the period from 7th July 2009 to 29th June 2011 for the S&P 500, NASDAQ and FTSE 100, 22th June 2009 to 29th June 2011 for the DAX 30 index and 23th June 2009 to 29th June 2011 for the CAC 40 index, with a total of 500 observations

**Figure 1: Levels of the S&P 500, NASDAQ, FTSE 100, DAX 30 and CAC 40 equity indices**

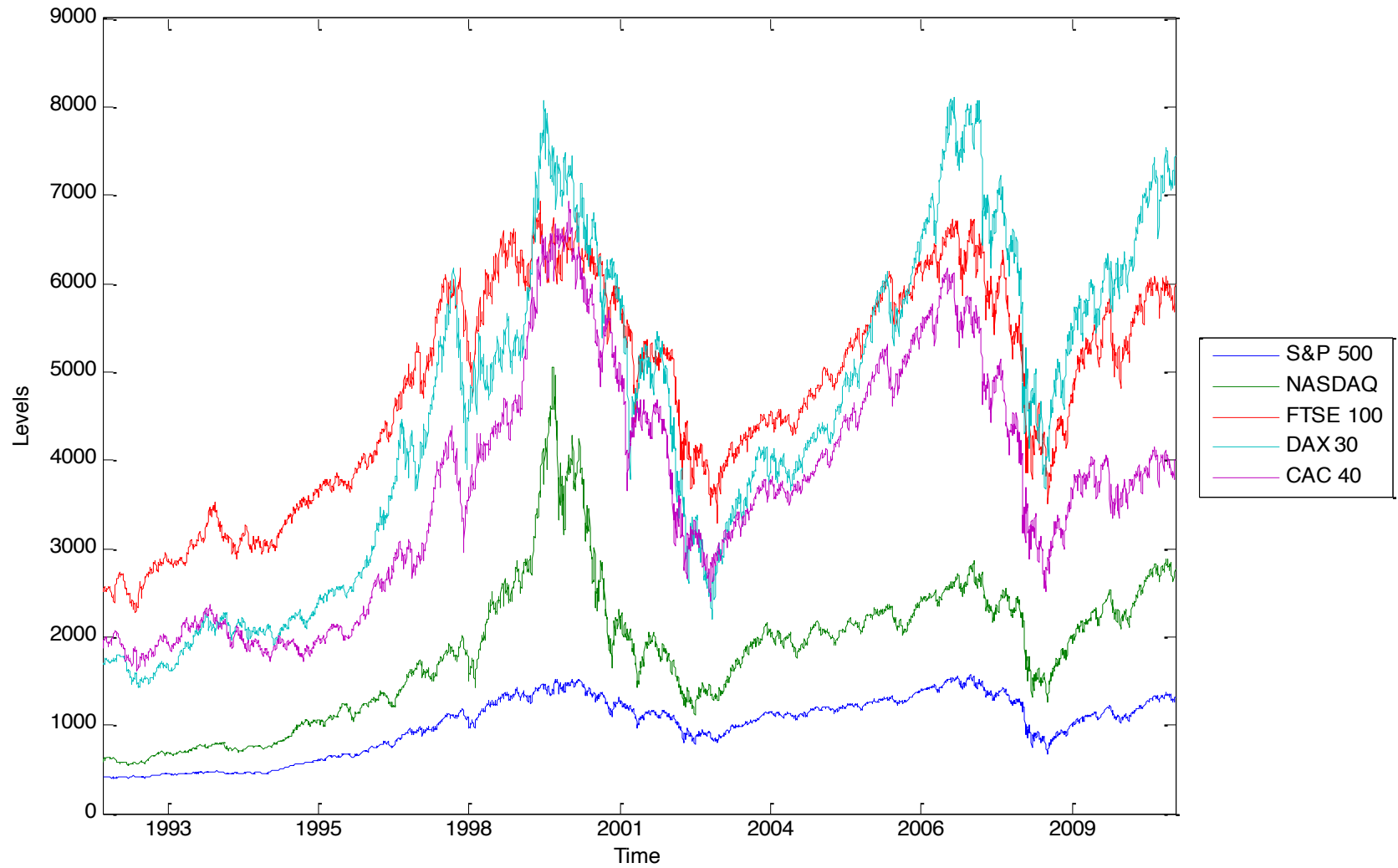

**Figure 2: In-sample volatility of the RiskMetrics$^{TM}$ with a decay factor of 0.94 of the S&P 500, NASDAQ, FTSE100, DAX30 and CAC40**

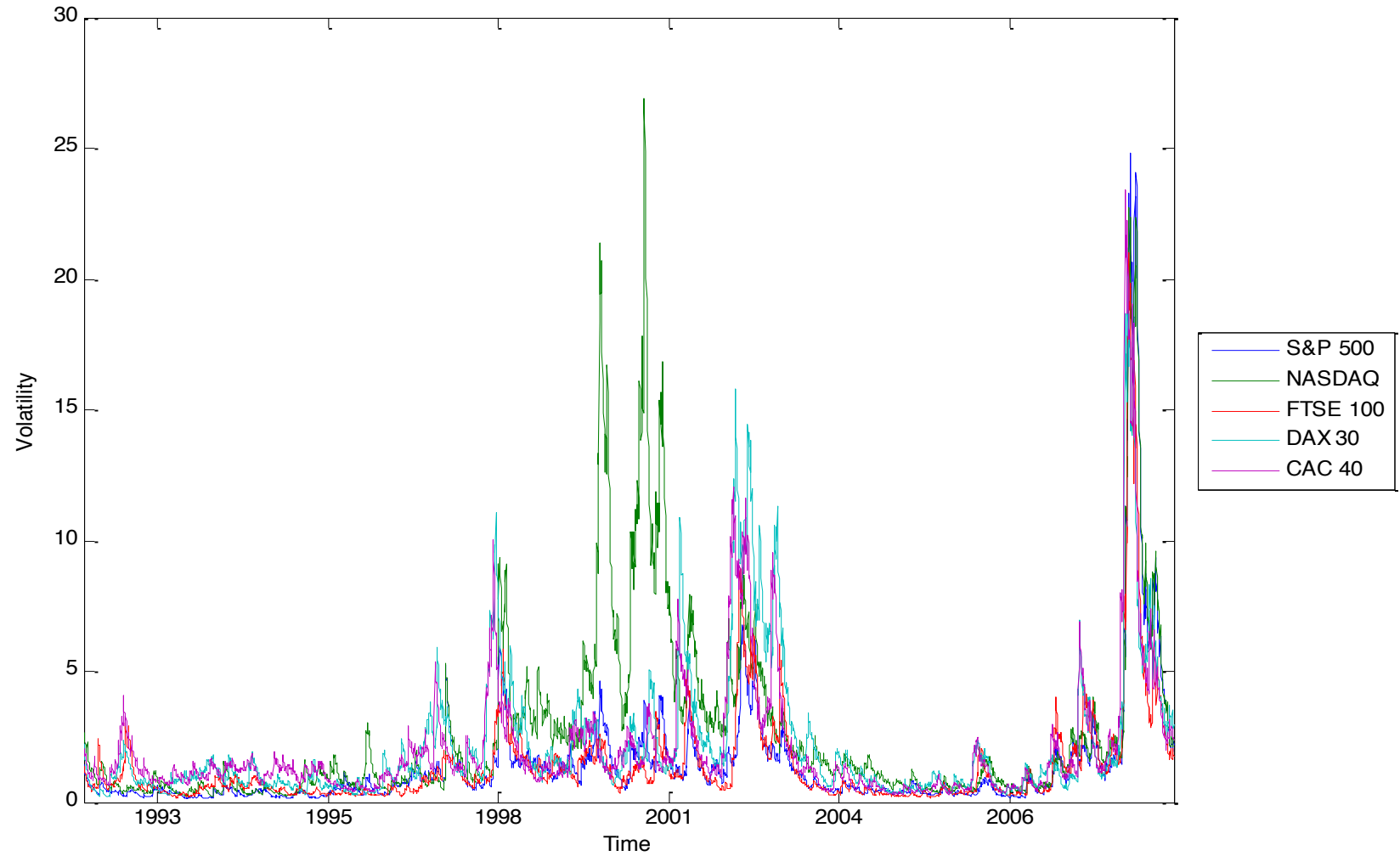

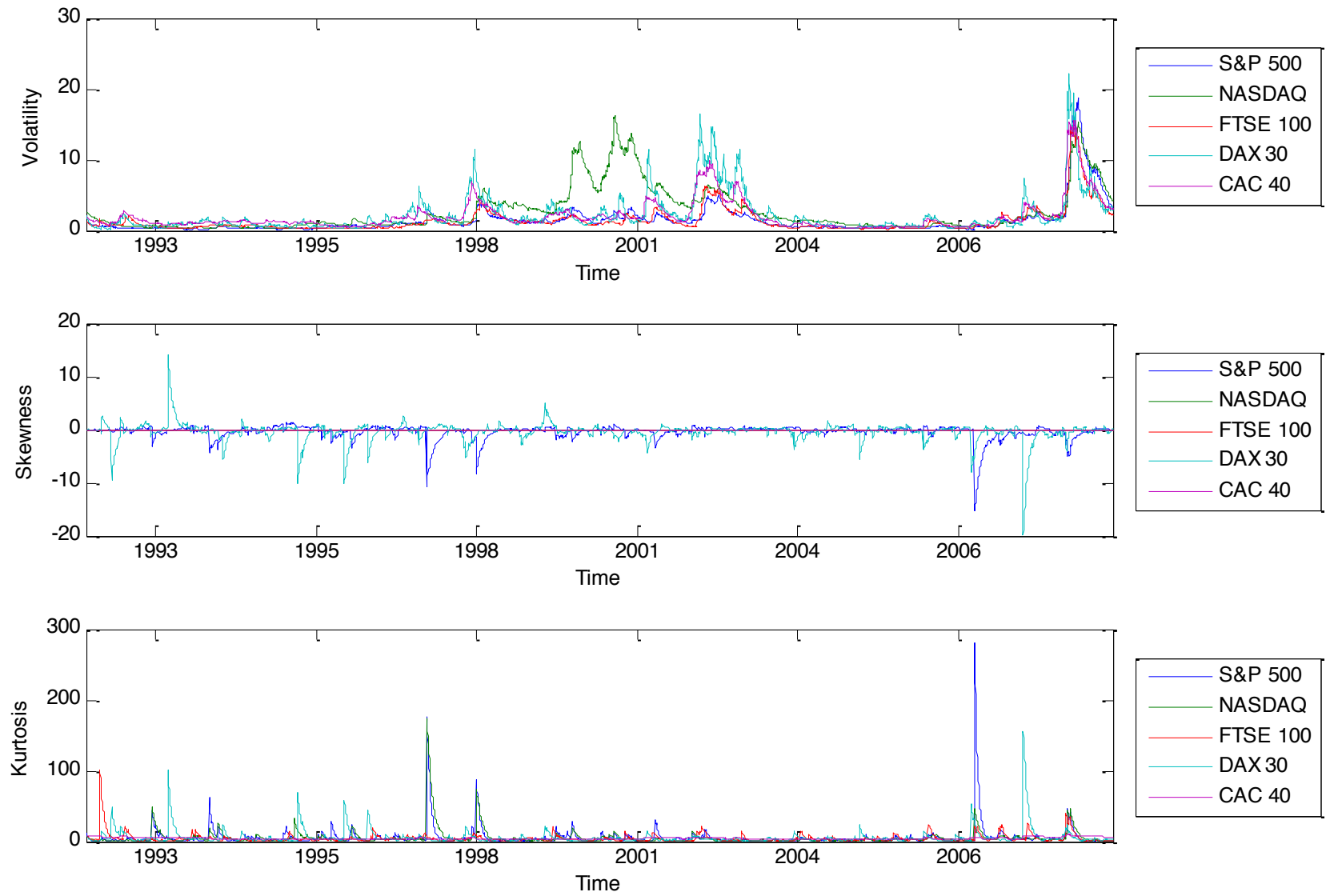

**Figure 3: In-sample volatility, skewness and kurtosis of the Modified Exponential Weighted Moving Average**

**Figure 4: In-sample volatility of the GARCH(1,1) model for the S&P 500, NASDAQ, FTSE 100, DAX 30 and CAC 40 equity indices**

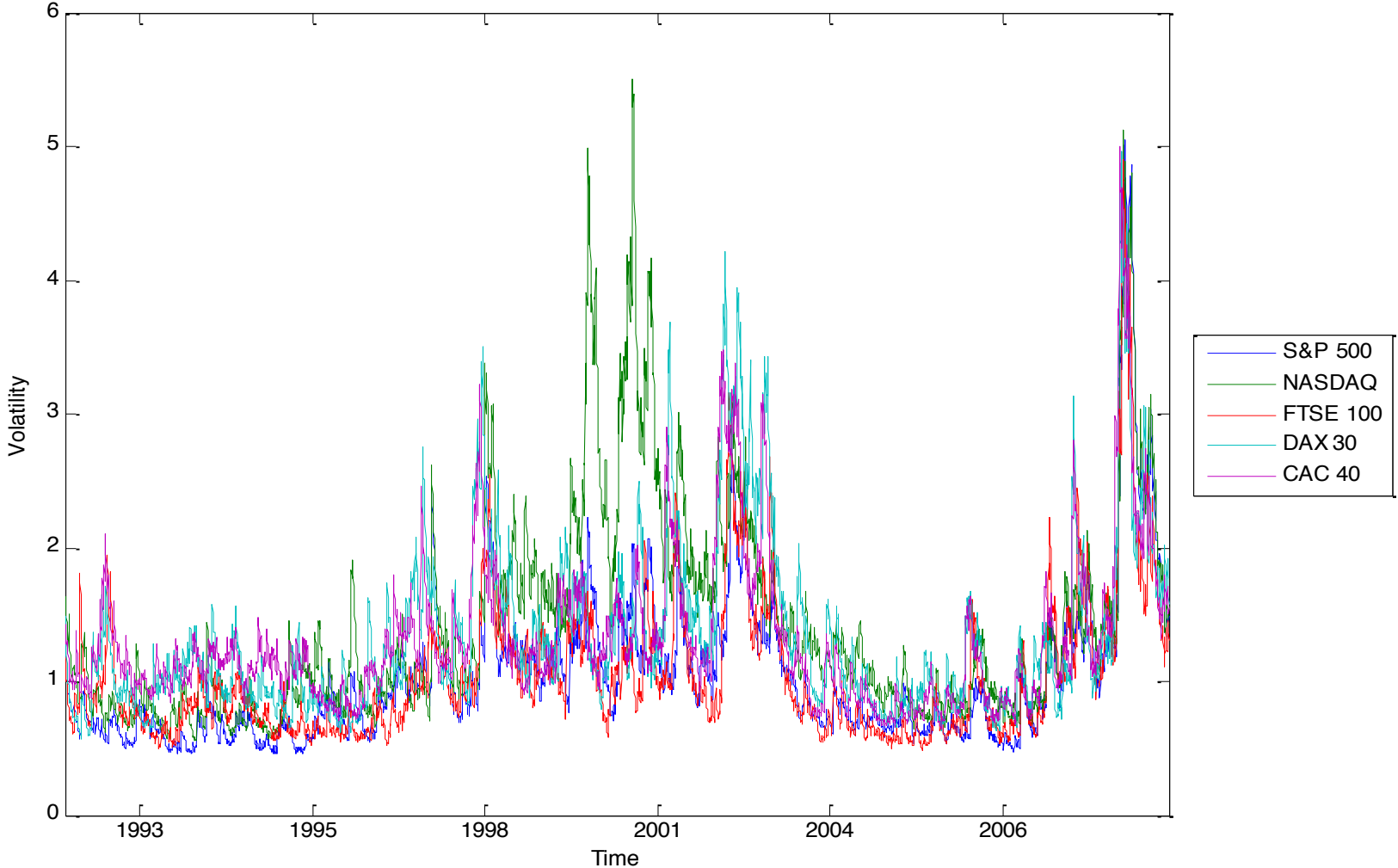

**Figure 5: Out-of-sample volatility forecasts and 99% VaR losses of the S&P 500 equity index**

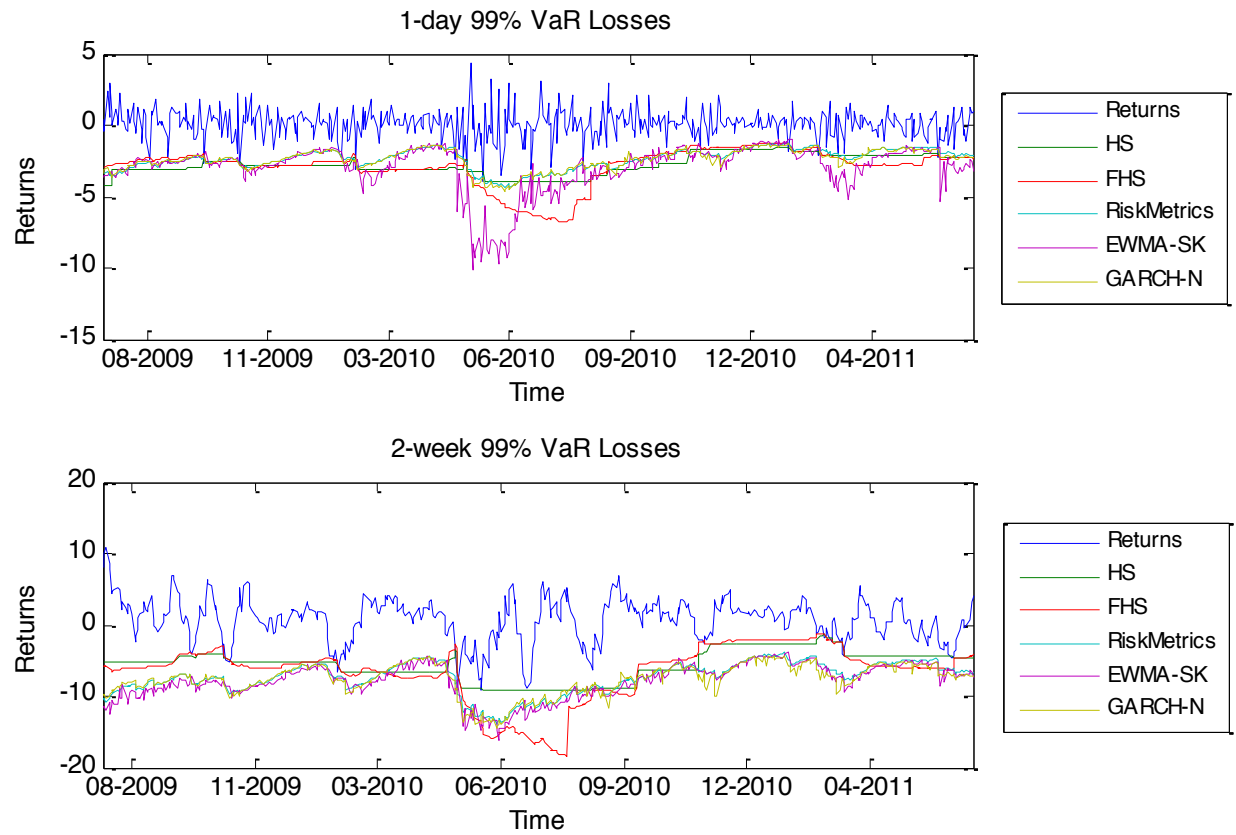

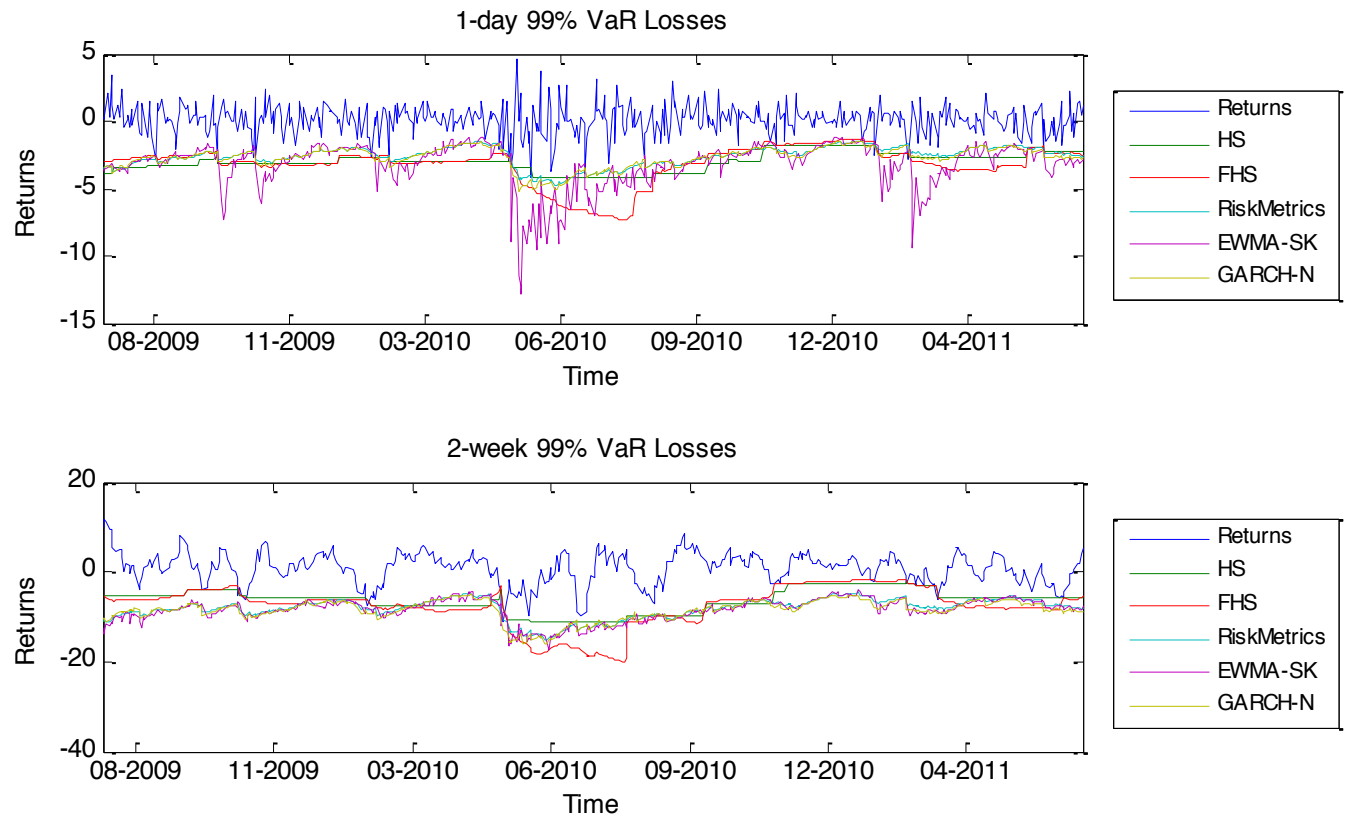

**Figure 6: Out-of-sample volatility forecasts and 99% VaR losses of the NASDAQ equity index**

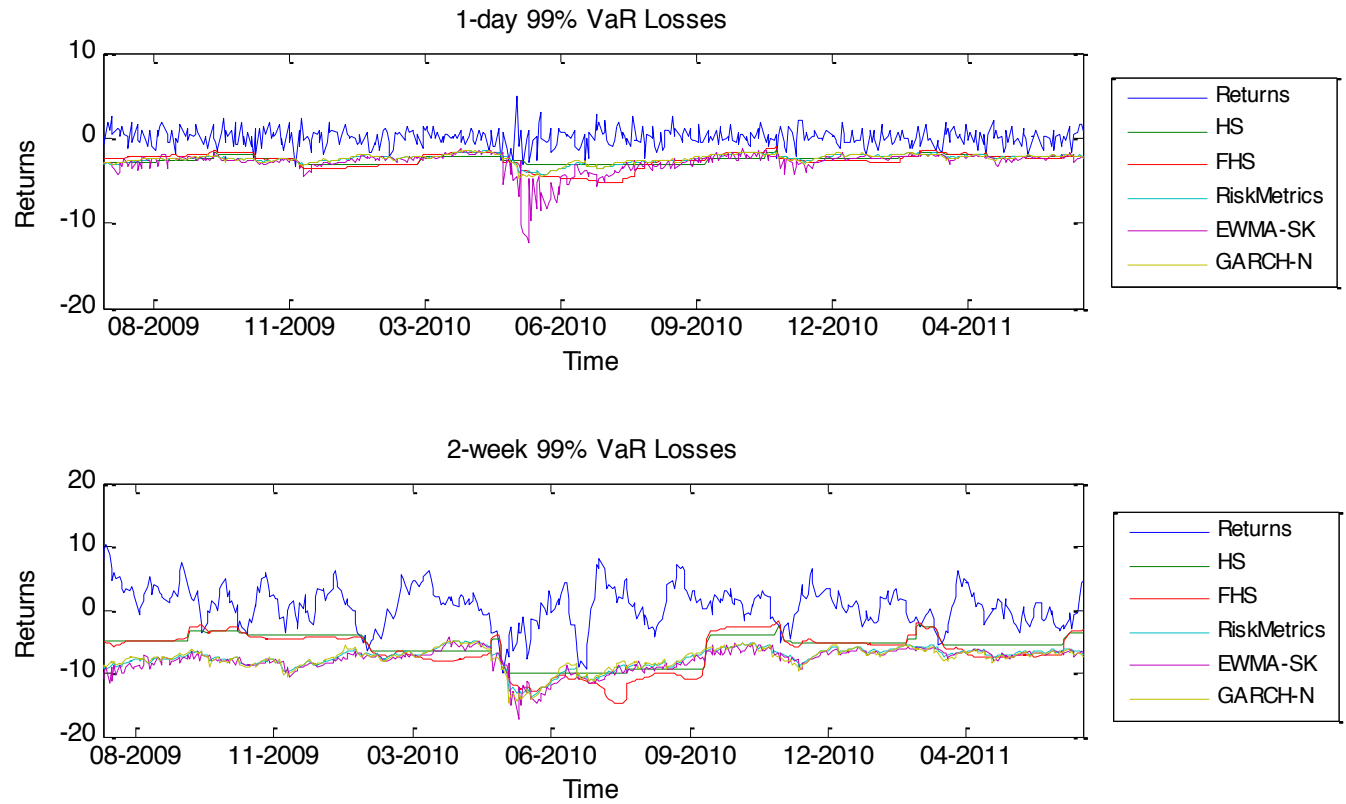

**Figure 7: Out-of-sample volatility forecasts and 99% VaR losses of the FTSE 100 equity index**

**Figure 8: Out-of-sample 1-day and 10-days volatility forecasts and 99% VaR losses of the DAX 30 equity index**

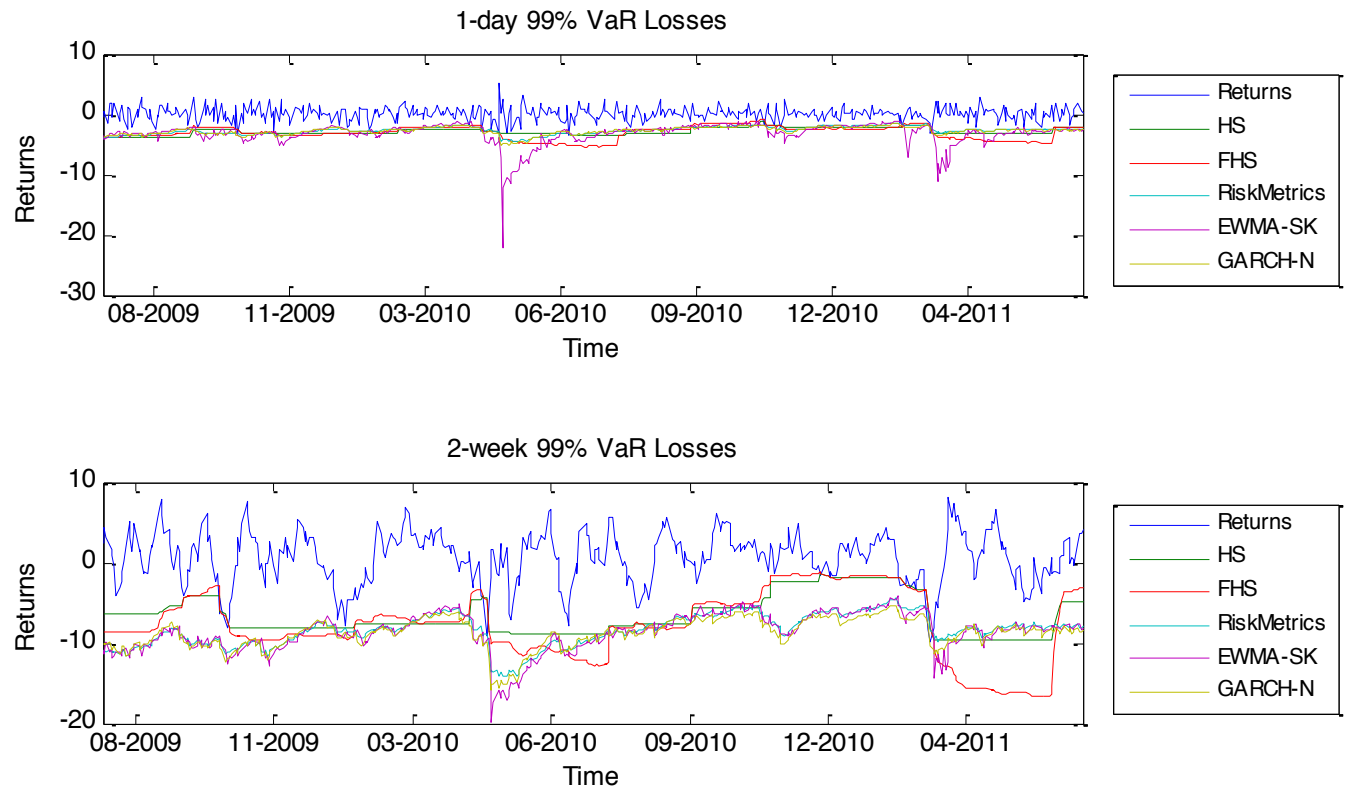

**Figure 9: Out-of-sample volatility forecasts and 99% VaR losses of the CAC 40 equity index**

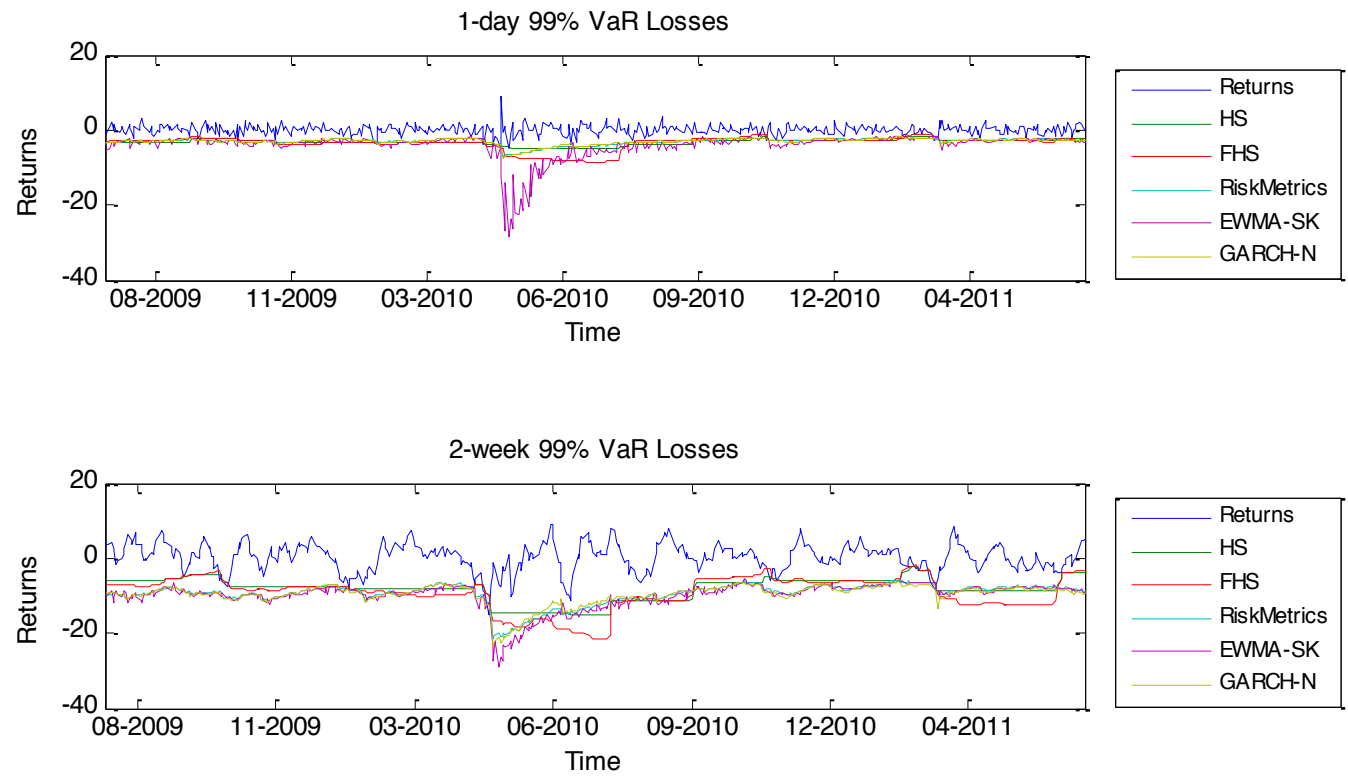

# APPENDIX A

## *Proof of the Galland and Tauchen (1989) proposed pdf that integrates to one*

This appendix illustrates that the function $f(\eta_t)$ in (8) integrates to one:

$$g(\eta_t) = \sum_{i=0}^{n} c_i He_i(\eta_t) \tag{A.1}$$

where $\{H_i(x)\}_{i \in N}$ represents the Hermite polynomials such that for $i \geq 2$ they hold the following recurrence relation:

$$H_i(x) = \left(xH_{i-1}(x) - \sqrt{i-1}H_{i-2}(x)\right)/\sqrt{i} \tag{A.2}$$

and they satisfy the following conditions:

$$\begin{aligned}
\int_{-\infty}^{\infty} H_i^2(x)\varphi(x)dx &= 1, \quad \forall i \\
\int_{-\infty}^{\infty} H_i(x)H_j(x)\varphi(x)dx &= 0, \quad \forall i \neq j
\end{aligned} \tag{A.3}$$

where $\varphi(\cdot)$ denotes the standard normal density function. The integration of the conditional density function given the condition in (A.3) becomes:

$$\begin{aligned}
&\frac{1}{\Gamma_t}\int_{-\infty}^{\infty} \varphi(\eta_t)\left[1 + \frac{s_t}{\sqrt{3!}}H_3(\eta_t) + \frac{k_t-3}{\sqrt{4!}}H_4(\eta_t)\right]^2 d\eta_t = \\
&= \frac{1}{\Gamma_t}\left[\int_{-\infty}^{\infty}\varphi(\eta_t)d\eta_t + \frac{s_t^2}{3!}\int_{-\infty}^{\infty}\varphi(\eta_t)H_3^2(\eta_t)d\eta_t + \frac{(k_t-3)^2}{4!}\int_{-\infty}^{\infty}\varphi(\eta_t)H_4^2(\eta_t)d\eta_t\right] = \\
&= \frac{1}{\Gamma_t}\left[1 + \frac{s_t^2}{3!} + \frac{(k_t-3)^2}{4!}\right] = 1
\end{aligned} \tag{A.4}$$

# APPENDIX B

## *Rescaling of the return distribution*

Given a return series of continuously compounded returns $r_t$ then the rescaled returns $R_x$ with the desired frequency, $x$, are computed as

$$R_x = \sum_{t=1}^{x} r_t \tag{B.1}$$

where the index $t$ denotes the time index (i.e. $t = 1, 2, \ldots, x$). Then the rescaled higher moments are estimated as

$$\mu_R = x\mu_r \tag{B.2}$$

$$\sigma_R = \sqrt{x}\sigma_r \tag{B.3}$$

$$s_R = \frac{s_r}{\sqrt{x}} \tag{B.4}$$

$$k_R = \frac{k_r + 3x(x-1)}{x} \tag{B.5}$$

For skewness it can be shown:

$$s_R = \frac{E(R - \mu_R)^3}{\sigma_R^3} = \frac{E\left(\sum_{t=1}^{x} r_t - x\mu_r\right)^3}{x\sqrt{x}\sigma_r^3} = $$

$$= \frac{\sum_{i=1}^{x}\sum_{j=1}^{x}\sum_{l=1}^{x} E\left[(r_i - \mu_r)(r_j - \mu_r)(r_l - \mu_r)\right]}{x\sqrt{x}\sigma_r^3} \tag{B.5}$$

$\because$

$$E\left[(r_i - \mu_r)(r_j - \mu_r)(r_l - \mu_r)\right] = \begin{cases} s_r\sigma_r^3, & \forall i = j = l \\ 0, & \forall i \neq j \neq l \end{cases} \tag{B.6}$$

$\therefore$

$$s_R = \frac{xs_r\sigma_r^3}{x\sqrt{x}\sigma_r^3} = \frac{s_r}{\sqrt{x}} \tag{B.7}$$

For kurtosis it can be shown:

$$k_R = \frac{E(R-\mu_R)^4}{\sigma_R^4} = \frac{E\left(\sum_{t=1}^{x} r_t - x\mu_r\right)^4}{x^2 \sigma_r^4} =$$

$$= \frac{\sum_{i=1}^{x}\sum_{j=1}^{x}\sum_{l=1}^{x}\sum_{k=1}^{x} E\left[(r_i - \mu_r)(r_j - \mu_r)(r_l - \mu_r)(r_k - \mu_r)\right]}{x^2 \sigma_r^4}$$

(B.8)

∵

$$E\left[(r_i - \mu_r)(r_j - \mu_r)(r_l - \mu_r)(r_k - \mu_r)\right] = \begin{cases} k_r \sigma_r^4, & \forall i = j = l = k \\ \sigma_r^4, & \forall \text{ respective two } i,j,k,l \text{ are same} \\ 0, & \forall i \neq j \neq l \neq k \end{cases}$$

(B.9)

∴

$$k_R = \frac{xk_r \sigma_r^4 + 3x(x-1)\sigma_r^4}{x^2 \sigma_r^4} = \frac{k_r + 3(x-1)}{x}$$

(B.10)